\documentstyle[prd,aps,twocolumn,floats]{revtex}
\begin{document} 
\twocolumn[\hsize\textwidth\columnwidth\hsize\csname
@twocolumnfalse\endcsname
\preprint{}
\title{The Case for Hyperbolic Theories of Dissipation 
in Relativistic Fluids}

\author{Angelo Marcello Anile\footnote{Electronic address:
anile@dipmat.unict.it}}
\address{Dipartimento di Matematica. Universit\'{a} 
degli Estudi di Catania. 95125 Catania. Italia}

\author{Diego Pav\'{on}\footnote{Electronic address:
diego@ulises.uab.es}}
\address{Departamento de F\'{\i}sica. Facultad de Ciencias.
Universidad Aut\'{o}noma de Barcelona. 08193 Bellaterra 
(Barcelona). Spain}

\author{Vittorio Romano\footnote{Email addres:
romano@dipmat.unict.it}}
\address{Politecnico di Bari. Sede di Taranto.
Viale del Turismo 8. 74100 Taranto. Italia}

\maketitle

\begin{abstract}
In this paper we highlight the fact that the physical content of
hyperbolic theories of relativistic dissipative fluids  is,
in general, much broader than that of the parabolic ones. This is
substantiated by presenting an ample range of dissipative 
fluids whose behavior noticeably departs from Navier-Stokes' .  
\\
\end{abstract}
\vspace{1.5cm}
]

\section{Introduction}
As is well-known causal theories of dissipative fluids, both
relativistic \cite{IS} and non-relativistic \cite{M}, \cite{JCL}
were set up
to get rid of some undesirable effects connected to the
conventional Eckart-type theories \cite{EL}. The advantages
of the causal theories are the following \cite{HL}: (i) For  
stable fluid configurations the dissipative signals propagate 
causally. (ii) Unlike Eckart-type's theories, there is no 
generic short-wavelength secular instability in causal 
theories. (iii) Even for rotating fluids, the perturbations 
have a well-posed initial value problem.

The central idea in causal-type theories (also called ``hyperbolic")
is to extend the space of variables of conventional theories
by incorporating in it the dissipative quantities (i.e., heat
fluxes, particle currents, shear and bulk stresses, etc.)
of the  phenomenon under consideration. 
Hence these quantities are                          
treated on the same footing as the ``conserved" variables (energy
density, particle numbers, etc.). The price to be paid
is a more involved theory with a larger number of variables
and parameters. In the approach of Israel [1.a-c] 
as well as in  the one by Pav\'{o}n {\em et al} [1.d], 
by combining the conservation of the stress-energy
tensor and the conservation of the particle current density
with the Gibbs equation (which has extra terms due to the enlargement 
of the space of variables) an expression is obtained for the entropy
production. The transport equations for the dissipative fluxes 
follow from the latter by demanding that the aforesaid production
be nonnegative. A distinguishing feature of these equations as  
compared with the corresponding ones in Eckart-type theories is
that the former contain a  ``relaxational term" -the convective time 
derivative of the corresponding flux multiplied by a parameter
with the dimension of time. Similar equations can be obtained 
in the extended thermodynamics approach of M\"{u}ller 
and Ruggeri [2.b]. 
The transport equations
can be understood as evolution equations for the 
dissipative variables as they (being ``hyperbolic" in nature)  
describe how these fluxes evolve
from an initial arbitrary state to a final steady one.
On the contrary, the Eckart-type theories 
are fit for quasi-steady states only, 
and they are usually referred to as ``parabolic" 
or ``quasi-stationary theories".

The aforesaid time parameter $\tau$ is usually interpreted  
as the relaxation time of the dissipative process. For instance, 
in heat conduction it is the time taken by the heat flux  to 
attain its steady value since the establishment of a  
temperature gradient across the system. It reflects the fact
that the effect (the heat flux or any other dissipative 
phenomenon) lags behind from its cause (the temperature gradient 
or any other thermodynamic ``force" acting on the system). 
This is directly linked to the ``second sound" 
-a well known phenomenon
in laboratory physics whose justification can be found 
within hyperbolic theories. We will turn to this topic below.
Even in the 
steady regime the descriptions offered by causal 
and acausal theories do not necessarily coincide. 
The differences between them in such a situation
arise from (i) the presence of $\tau$ in terms that 
couple the vorticity to the heat flux and shear stresses.
These may be large even in steady states (e.g. rotating
stars). There are also other acceleration coupling terms
to bulk and shear stresses and heat flux -see for instance
[5.a]. The coefficients
for these vanish in parabolic theories, and they could
be large even in the steady state. (ii) From the convective 
part of the time derivative (which are not negligible 
in the presence of large spatial gradients). (iii) From 
modifications in the equations of state due to 
the presence of dissipative fluxes. However, it is precisely before
the establishment of the steady regime that both types of 
theories (hyperbolic and parabolic) can differ more importantly.
Therefore if one wishes to study a dissipative process for 
times shorter than $\tau$, it is mandatory to resort to a hyperbolic 
theory which is a more accurate macroscopic approximation
to the underlying kinetic description.
Only for times longer than $\tau$ it is permissible
to go to a parabolic one (provided that the spatial
gradients are not so large that the convective part of the time
derivative does not become important, and that the fluxes
and coupling terms remain safely small).
All this has long been well-known.

Recently, however, it has been argued that the physical
content of causal (hyperbolic) and stationary (parabolic)
theories {\it is the same} \cite{RG}, \cite{LL}. Such a claim 
is based on the study of simple dissipative fluids  
under the assumption that its relaxation time is of the same 
order than the colision time between particles -something 
that, as we shall show is far from being always the case. 
Moreover in \cite{LL} it is asserted that such a study is 
{\it very general and should apply to essentially any 
theory of fluids (including those describing superfluids, 
mixtures of different kinds of fluids, etc.)}; and that
{\it the meaningful differences between the causal theories
and the non-causal Navier-Stokes theory can not be observed.
The complicated dynamical structure of the causal theories 
is necessary to insure that the fluid evolves in a causal 
and stable way. But this rich dynamical structure is 
unobservable, since the physical states of a fluid always 
evolve in a way that is also well described  by the 
Navier-Stokes expressions for the stress-energy tensor, etc.}
These assertions (as well as some others of the like in 
\cite{RG} and \cite{LL}) are misleading (to say the least) 
and likely to promt confusion.
The purpose of \cite{RG} was to justify some class of
relativistic,
hyperbolic theories of dissipative fluids. This was 
achieved for simple fluids under the assumptions that
(i) the departure of these fluids from the 
Fourier-Navier-Stokes behavior is negligible small, 
and (ii) the relaxation times are extremly short on
macroscopic scales. Notice that in \cite{RG} it is not 
explictly denied that on examining a typical FNS
(Fourier-Navier-Stokes) fluid
in certain regimes, or with sufficient accuracy, there
will be found non-FNS behavior, which could be 
satisfactorily described  by some hyperbolic theory. 
However, it is  explicitly suggested that {\it it may 
happen that the hyperbolic theories have no useful physical 
consequences at all}. 

The aim of the present paper is to 
explicitly show that the physical content of hyperbolic
(causal) transport theories is much broader than the 
corresponding one of parabolic (quasi-stationary) theories.
We will do this by presenting a series of well-known
physical results in an ample range of dissipative systems
including simple fluids. A criticism 
in \cite{RG} and \cite{LL} of the hyperbolic theories 
is that they contain a number of 
{\it arbitrary} parameters.
We would like to emphasize that this is not the case. 
In fact these parameters can be determined
provided the state equations of the fluid
under consideration are known. From these such parameters 
can be calculated either by kinetic theory or fluctuation 
theory. Israel and Stewart computed them for a quantum gas
using Grad's 14-moment approximation -see the set of
equations (7.8) in [1.b]. 
For an example in radiative fluids, in which the 
coefficients connecting the heat flux with the gradient
of the dissipative stresses are computed, see \cite{JD}. 
Further, a  clear conceptual difference 
between the physical content
of both set of theories is the following one. With the  
help of Einstein's theory of thermodynamic fluctuations  
hyperbolic theories are capable of deriving the transport 
coefficients (heat conductivity $\kappa$, bulk $\zeta$
and shear $\eta$ stresses, etc) from the fluctuations,
because they are based on an extended entropy (which
comprises the dissipative fluxes), at variance 
with parabolic theories. This has been done,
for instance, for radiative fluids \cite{PJC}. This
however cannot be done in the case of parabolic theories,
which necessarily have to be supplemented with microscopic
expressions to obtain these transport coefficients.

Section II recalls the concept of relaxation time. Section
III brings about seven examples in the non-relativistic 
context showing that the physical content of 
hyperbolic theories is much wider than that of the parabolic
ones. These include examples involving heat and/or electric
conduction in solids. Note that there is nothing wrong in 
that because it is the electron and the phonon gases in 
the solid that are responsible for such
transport processes within it \cite{ASS}. Thereby it is 
meaniningful to apply, in this regard, fluid theories 
to solids. Section IV deals with four fully relativistic
examples. In Section V our main conclusions and outlooks
are summarized.  
\section{The relaxation time}
Before going any further it is expedient to say a few words 
about $\tau$. The relaxation time is a key quantity 
in hyperbolic systems with a distinct physical meaning
as made explicit before. It is connected to the mean 
collisional time of the fluid particles $t_{col}$,
sometimes erroneously identified with it. 
In principle they are different since $\tau$ is a 
macroscopic time, although in some instances 
it may correspond to a few $t_{col}$. It is more 
appropriate
to interpret $\tau$ as the time taken by the 
corresponding dissipative flux to relax to its
steady value. We believe that at the root of the 
confusion raised by \cite{RG} and \cite{LL}
is, among other things,  the assumption that 
$\tau$ and $t_{col}$ are {\it always}
of the same order. In fact one of the main 
arguments in Ref \cite{LL} heavily rests on the 
assumption that the dimensionless quantity
$\Gamma \equiv (\tau c_{s}/L)^{2}$ 
is always negligible. Likewise it is crucially 
needed for the bounds established by equations (3.22) and 
(3.23) in \cite{RG} to hold. (Here $c_{s}$ stands for the 
isentropic speed of sound in the fluid under
consideration and $L$ the characteristic length of  the 
system). That assumption would be right
if $\tau$ were always comparable to
$t_{col}$ and $L$ always ``large", 
but as we shall see there are 
important situations in which 
$\tau \gg t_{col}$, and  $L$ ``small" 
although still large enough to justify 
a macroscopic description.
For tiny semiconductor pieces of about
$10^{-4}$ cm in size, used in common electronic
devices submitted to high electric fields, the
above dimensionless combination (with 
$\tau \sim 10^{-10}$ sec, $c_{s} \sim 10^{7}$ cm/sec
\cite{AMM}) can easily be of the order of unity.
In ultrasound propagation as well as light-scattering 
experiments in gases 
and  neutron-scattering in liquids
the relevant length is no longer the system size, 
but the wavelenght $\lambda$ which is usually much
smaller than $L$ \cite{JCL}, \cite {WM},\cite{JRD}. 
Because of this, 
hyperbolic theories may bear some importance
in the study of nanoparticles and quantum dots.

Likewise in polymeric fluids relaxation 
times are related to the internal configurational
degres of freedom and so much longer than $t_{col}$
(in fact they are in the range of the minutes), and 
$c_{s} \sim 10^{5}$ cm/sec, 
hence it follows  $\Gamma \sim {\cal O}(1)$.
In the degenerate core of aged stars the 
thermal relaxation time can be as high as $1$ second
\cite{MH}. Assuming the radius of the core of about
$10^{-2} R_{SUN}$ one has 
$\Gamma \sim {\cal O}(1)$ again. 
Think for instance of  
some syrup fluid flowing under a shear stress, 
and imagine 
that the imposed shear is suddenly suppressed.
This liquid will come to rest only 
after a much longer time
($\tau$) than the collision time between its constituent
particles has elapsed. Many other examples could be added
but we do not mean to be exhaustive.

In transport theory when the mean free-path 
of the particles is much smaller
than the macroscopic length scale of interest, 
the gas is in a state of small
deviations from local thermal equilibrium (LTE) 
and the usual continuum
assumptions are appropriate. 
In the gas dynamical case the local thermal
equilibrium equations are the FNS 
system which are parabolic in nature. When the mean
free-path is of order of the macroscopic length 
scale the usual continuum
description is no longer valid and the 
more fundamental kinetic approach
must be considered. Under rarefied situations 
(i.e., limiting to the one particle distribution 
function) the kinetic description is embodied in the
Boltzmann Transport Equation (BTE). Solution techniques 
for this equation fall in two main groups: 
(i) direct, particle methods, and  
(ii) extended continuum approaches 
(e.g. based on moment methods). In general, direct
particle methods (e.g. Monte Carlo) are capable of 
obtaining solutions under general conditions, 
whereas extended continuum approaches tend to
be restricted to some definite ranges of the parameters. 
However, for transitional fluxes ranging from very large 
mean free-paths (where only a direct particle method could 
be applicable) and very small mean free-path
(where a usual continuum description is warranted) 
to utilize direct particle methods would lead 
to prohibitive computational costs. Under
these conditions extended continuum methods might 
be computationally much
more efficient. In the extended thermodynamics 
framework the moment equations
obtained from the Boltzmann transport 
equation are closed by utilizing
the entropy principle (M\"{u}ller {\em et al}
[2.b]; 
Jou {\em et al} \cite{JCL}) or by
assuming the maximum entropy distribution function 
(i.e., a special form for
the distribution function which maximizes 
entropy subjected to the given moments \cite{JCL}). 
The resulting set of transport equations form a quasi-linear,
symmetric and hyperbolic system of conservation laws 
(in a neighbourhood of LTE) 
which admits a supplementary conservation 
law in the form of entropy balance. From the mathematical 
viewpoint this system has desirable properties.
The obtained mathematical model is able to describe 
efficiently the regimes that lie in between 
the collisionless and the collisional regimes, i.e.,
the transitional one.  In fact these models capture 
the usual continuum regime when the mean free-path 
is much smaller than the macroscopic length
scale while in the transition regime provide 
values for the physical
quantities that are consistent with a 
positive definite distribution
function, and therefore are physically 
more satisfactory. These models are
potentially very useful for constructing 
hybrid continuum/kinetic schemes for simulation 
purposes (i.e., numerical schemes where the kinetic
approach, which is computationally very expensive, 
is used only where it is
required, that is, in the collisionless case, 
where continuum equations are
used in the near LTE, and extended continuum 
models in the transition regions)
\cite{LEV}. Therefore, hyperbolic theories broaden
the domain of continuum models and bridge the gap
between collisional and collisonless regimes. Then 
the scope of hyperbolic theories is to describe the
transitional regimes and it is meaningless to 
restrict them just to the collision dominated case.
\section{Non-relativistic examples}
\begin{enumerate}
\item {\it Heat flux and shear stress in ideal gases}\\
Let us consider an ideal gas subjected 
to a temperature gradient and a shear stress. 
It has been established  
with the help of Grad's thirteen moment approximation
\cite{GD} that the the relaxation times for the heat flux
and the shear pressure are given by
$\tau_{q} = 3/(2nm \gamma)$ and 
$\tau_{s} = 2 \tau_{q}/3$, respectively. 
Here $n$ stands for the particle number density,
$m$ the molecular mass, and $\gamma$ a positive
constant related to the interaction 
cross section -see \S3.3 in Ref \cite{JCL}. 
It is obvious that in a collision dominated regime
$n$ will be large and hence both relaxation times 
will not be much longer than $t_{col}$. In the
opposite extreme (Knudsen regime) 
the continuum approximation breaks down.
However, for intermediate situations (where hyperbolic
theories apply) both $\tau_{q}$ and $\tau_{s}$
can be substantially larger than $t_{col}$.

\item{\it Second sound in superfluids and solids}\\
The second sound  was first discovered by
Peshkov \cite{PESH} in Helium II in the mid-forties, 
and later confirmed, also in solids,
by a number of experimentalists (see,
e.g. \cite{TZOU}, \cite{RAG} and \cite{WDH}). 
It is related to the 
propagation of the heat signal in solids 
not at the phase speed predicted by Fourier's law,
\[ v_{ph} = \sqrt{2 \chi \omega} \; \] 
(which diverges as the frequency of the signal 
$\omega \rightarrow \infty$),
with 
\[ \chi \equiv \kappa /(\rho c_{v}) \, , \]
where
$\kappa$, $\rho$ and $c_{v}$ stand for the
thermal heat conductivity, the mass density and the 
specific heat at constant volume of the solid, respectively,
but at the speed
\[ 
v_{ph} = \frac{\sqrt{2 \chi \omega}}{\sqrt{ \tau \omega 
+ \sqrt{1 + \tau^2 \omega^2}}},
\]  
which remains finite at high frequencies.
This result straightforwardly follows from combining 
the Maxwell-Cattaneo equation,
\[ 
\vec{q} + \tau \frac{ \partial \vec{q}}{ \partial t} = 
- \kappa \nabla T \; ,
\]
typical of hyperbolic theories, 
with the equation for the balance of the 
energy in a rigid solid,
\[ \rho \dot{\epsilon} = - \nabla \cdot{\vec{q}}\;  \]
-see \S6.1 of \cite{JCL} for details. At low frequencies
(i.e., $\tau \omega \ll 1$) the phase velocity of 
the hyperbolic theory reduces to the one predicted by 
the Fourier's law. However, outside that range both
equations significantly differ from one another; and in the
high frequency limit ($\tau \omega \gg 1$) the former
reduces to \[ U = \sqrt{ \chi/ \tau} \; , \] 
which is the propagation
speed of thermal pulses -see \S 6.2 in \cite{JCL}.
As mentioned above all these results have been 
confirmed experimentally  -see \cite{WDH}
and  \cite{DDJ} for a review.  
Notice that if one makes the unphysical assumption
implied by equating $\tau$ to zero (which corresponds to
the Fourier's law) the latter expression diverges.

Closely connected to the above discussion is the 
response of material sample to an instantaneous 
heat pulse (activated by means of a laser beam, 
or an electric discharge). An elegant treatment
of the problem  
can be implemented by adding a 
delta-like term to the right hand side of 
the energy balance equation. This when 
combined with the Maxwell-Cattaneo equation
leads to a second order differential equation 
in terms of the associated Green function, 
and one obtains for the temperature response
to the heat pulse the expression
\begin{eqnarray}
\Delta T(x, t) =  \frac{g_{0}}{4 \kappa} \exp 
\left(-\frac{t}{2\tau} \right)
\left[2 I_{0}(\xi) U \tau \delta(Ut- \mid x \mid) \phantom{\frac{1}{2}}\ \ 
\right.  
&&\nonumber\\
+\left. \left( I_{0}(\xi) + \frac{t}{\tau \xi} I_{1}(\xi) \right)
H(Ut - \mid x \mid ) \right] && \; ,
\nonumber
\end{eqnarray}
(where, for simplicity, we have restricted ourselves 
to one spatial dimension)
while the temperature response corresponding to the 
parabolic theory (Fourier's law) is 
\[
\Delta T(x, t) = \frac{g_{0}}{2\sqrt{\pi \chi t}} 
\exp \left(- \frac{x^2}{4 \chi t} \right) \; .
\]
Here $g_{0}$ is the energy supplied to the sample,
$\xi \equiv \frac{1}{2} \sqrt{(t/\tau)^2 -(x/U \tau)^2}$,
$I_{0}(\xi)$ and $I_{1}(\xi)$ are related to the 
Bessel functions of the $n$ kind by
$I_n(\xi)= -(-i)^{n} J_{n}(i \xi)$,
and $H(U -t \mid x \mid)$ is the Heaviside unit step
function.

There are important differences between the predictions
of the conventional parabolic theory and the causal
theory. (i) In the former, the response to a heat pulse
is instantaneously felt in all the sample. In the latter,
by contrast, it vanishes for $ x > U t$. (ii) Two distinct 
contributions enter $\Delta T_{hyperbolic}$. On the one 
hand there is the term with $\delta (Ut - x)$ 
which reflects the initial pulse damped by 
an exponential factor; on the 
other hand there is, in addition, 
the Heaviside term which corresponds to the
wake. The latter after a sufficiently long time reduces
to the conventional diffusion solution. Such a structure 
has been observed in the propagation of heat pulses 
in solids at low temperatures (see
Figure 1 for a schematic representation of the 
experimental results) and in the structure of X-ray peaks
from X-ray bursters in star explosions -see point 1 
of Section IV below. (iii) The hyperbolic description
predicts the presence of a precursor peak which has
no counterpart in the parabolic theory. Although both 
situations are characterized by the same total energy,
the distribution of the thermal energy differs from
one another. While in the acausal description the
energy is roughly evenly distributed in the whole
spatial region, in the causal description, 
by contrast, there is a sharp concentrantion 
of energy associated to the precursor peak.
\item {\it Ultrasond waves and light-scattering
in ideal monoatomic gases}\\
In the low frequency regime (i.e., $\tau \omega \ll 1$)
the sound propagation as well as light-scattering
in ideal gases, with a heat flux and
shear stress, are well described by the parabolic FNS 
theory of dissipative phenomena. However, as $\omega$
is increased the limitations of this theory
begin to be felt 
(in fact as soon as $\tau \omega \simeq 0.5$ notorious 
discrepancies with respect to the 
FNS theory appear); 
and in the high frequency regime
($\tau \omega \gg 1$) the disagreements 
with experimental results become wider.
Aside from predicting  an unbounded sound speed as
$\omega \rightarrow \infty$ it fails to foretell
the existence of an additional wave and
yields erroneous values for the absorption 
coefficient. Things fare differently for 
hyperbolic theories. These predict 
constant phase speed for both waves, 
$1.64 \, c_{s}$ and $0.72 \, c_{s}$ with
$c_{s} = \sqrt{5 \, k_{B}T/(3m)}$, 
as $\omega \rightarrow \infty$
\cite{AP}, \cite{LC}. All these results
show good agreement with the well-known 
experimental data of Meyer and Sessler
\cite{MS} and Greenspan \cite{MG} about
ultrasound waves.

Table I contrasts the predictions of the FNS 
and hyperbolic theories \cite{CM}
with the corresponding experimental results in terms of the ratio
between the mean free path $l$ of the moleculas and the wavelength
$\lambda$ of the soundwave.

Light scattering in dilute ideal monoatomic gases has been
studied in detail by Weiss and M\"{u}ller \cite{WM}. 
These authors find good agreement of the hyperbolic theory
with the experimental data of Clark \cite{CLARK} (see figures
7.1 through 7.3 for the dynamic form factor in \cite{WM})
while the parabolic FNS theory fails hopelessly.  

\item {\it Electric circuits}\\
Simpler and more familiar examples
can be found in the 
province of electric circuits. To fix ideas consider
a resistor and an inductance connected to a
source ${\cal E} (t)$. The intensity in the resulting 
circuit is governed by
${\cal E} (t) = R I(t) + L \, dI(t)/dt $. 
Note that if the source 
were suddenly turned off, the current will decay in 
a characteristic (relaxation) time given by
$\tau_{e} = L/R $. It is apparent that by 
suitably choosing $R$ and $L$ the aforesaid 
time can be made much longer than the collisional 
time of the electrons in the circuit.
\item {\it Thermohaline instability}\\
Imagine a layer of warm salt water above a layer of fresh
cold water. This system will be dynamically stable against
sinking so long the warmth of the salt water is high enough
to reduce its specific weight to below that of the fresh
water. However, as the upper layer cools off, its density
augments and eventually small blobs of salty water will sink
down. This kind of instability may also occur in stars \cite{KR}. 
Let $\Delta T$ be the temperature difference between a generic blob
of salt water and its surrounding and 
\[
\tau_{d} = \frac{ \chi \rho^2 a^2 c_{p}}{4 a_{B} c T^3} \; \; 
\]
the adjustment time where $\chi$ 
denotes the mean absorption coeffcient, $\rho$ the mass density 
of the blob,
$a$ the radius assuming the blob spherical, 
$a_{B}$ the Boltzmann radiation constant,
and $c_{p}$ the specific heat at constant pressure. If the 
evolution of $\Delta T$ is analyzed using the Fourier law 
for the the conductive heat transport
$\vec{q} = -{\kappa} \nabla T$,
one obtains an exponential decrease
$\Delta T = \Delta T(t = 0) exp(-t/\tau_{d})$, 
and the blob simply sinks in the water below it.
However, if the Maxwell-Cattaneo equation
is used in place of Fourier's one, it follows
\begin{eqnarray}
\Delta T(t) = \Delta T(t = 0) e^{-x} \left\{
\cos(\omega x) \phantom{\frac{1}{2}} \right. &&\nonumber\\ 
- \left. \left( 1 + \frac{2 \beta}
{\Delta T(0) \tau_{d}} \right) \frac{ \sin(\omega x)}
{\omega} \right\} &&
\nonumber
\end{eqnarray}
with $ x \equiv t/2 \tau$, 
$\omega \equiv \sqrt{ 4 \tau/\tau_{d} - 1}$ and  
\[
\beta \equiv \int_{-\infty}^{0} \Delta T(t) exp(t/\tau) dt
\]
and the blob undergoes an oscillatory motion
-see \cite{HF} for details. So for times shorter than
the effective time for relaxation into diffusion 
the behavior of the blobs differ drastically from 
that predicted by the parabolic theory.
\item {\it Polymeric fluids}\\
Simple models of dilute polymeric solutions accounting nicely
for the observed dynamics of these fluids are, among others,
those of Rouse and Zimm \cite{BAH}. In these the polymer 
is assimilated to a chain of beads connected by 
Hookean springs of constant $K$, and 
the viscous pressure tensor corresponding to 
the mode $m$ is governed by the Maxwell-Cattaeno type
equation
\[
P^{(m)}_{ik} + \tau^{(m)} \frac{\partial P^{(m)}_{ik}}
{\partial t} = - \frac{1}{2} \eta^{(m)} \left( 
\frac{\partial v_{i}}{\partial x_{k}} + 
\frac{\partial v_{k}}{\partial x_{i}} \right)
\]
with $\eta^{(m)}$ the viscosity coefficient and 
$\tau^{(m)} = \xi/(2K a^{(m)})$
the relaxation time of that mode. Here $\xi$ 
is the friction coefficient, 
$a^{(m)} = 4 \sin (\pi m/2N)$ 
the eigenvalues of the Rouse matrices
and $N$ the number of monomers. It is apparent that
$\tau^{(m)}$ can be made as large as desired 
just by increasing this number. For large $N$
one has $\tau^{(m)} \sim N^{2}$.
The above equation 
for $P^{(m)}_{ik}$ can be easily derived in the 
framework of hyperbolic theories - see for instance
Chapter 7 in Ref \cite{JCL}. 
\item {\it Carrier transport in semiconductor devices}\\
Enhanced functional integration in modern electron devices requires
an accurate modelling of charge carrier transport in doped semiconductors
in order to describe high field phenomena such as hot electrons,
impact ionization, etc \cite{HAE}. The standard drift-diffusion
equations \cite{SEL} cannot describe adequately high field and
submicron phenomena because they hold only in the limit of weak
electric field and short mean free-paths \cite{MRS}.
The ultimate description (at least at the semiclassical level) is to
find a direct solution to the semiclassical BTE \cite{MRS} 
by numerical techniques such as
Monte Carlo simulation \cite{FLX}. Such an
approach, however, requires very long computing times ad is therefore
unsuitable for engineering design applications. An intermediate approach
is to write a relatively small number of moment equations of the BTE,
closing the system of equations by physical and mathematical ansatz
(the closure assumption) which must then be checked by Monte Carlo
simulation. The obtained macroscopic models, with the transport 
parameters (relaxation times, heat conduction coefficient, etc.)
obtained by Monte Carlo verified physical and mathematical ansatz, are
called hydrodynamical models and their range of validity is somehow in
between the drift-diffusion regime and the kinetic one.\\
Lately Anile {\em et al} \cite{APT}, \cite {AMM}
have introduced an extended hydrodynamical model
in which the closure of the moment equations is achieved by imposing the
entropy principle of extended thermodynamics [2.b]. In the
case of silicon semiconductors, there are several distinct time scales,
the energy relaxation time  $\tau_{w}$, the momentum relaxation time
$\tau_{p}$, the energy flux relaxation time $\tau_{q}$, etc. Monte 
Carlo simulation shows that  $\tau_{w} >>\tau_{p}>\tau_{q}$, 
due to the predominance of elastic collisions, 
and this justifies the existence of a local thermal 
equilibrium state in which the electron temperature 
is much larger than the lattice temperature,
and the electron gas average momentum and energy flux vanish. 
In the extended hydrodynamical model of Anile and Muscato
\cite{AMM}, the first thirteen moment 
equations are considered and the closure,
at the level of the flux of energy flux, 
is achieved by imposing compliance
with the entropy principle up to second order around the state of local
thermal equilibrium (supplemented by the maximum entropy ansatz for the
distribution function in order to calculate one of the coefficients in the
expansion, which otherwise would have remained undetermined
\cite{AU}, \cite{AAT}). This
closure has been checked by Monte Carlo simulation in several 
cases.
The resulting system of equations is hyperbolic in a broad domain of 
variables. In particular the heat flux obeys an evolution equation of
the Maxwell-Cattaneo type (and likewise the viscous stresses, which 
however are unimportant in this context and will be ignored).
In the case of small gradients (of temperature, velocity, etc.) one can 
linearize the extended system around the state of local thermal equilibrium
(the so-called Maxwellian iteration \cite{MUN}), thereby
obtaining a constitutive law for the heat flux . It turns out that the
constitutive law for the heat flux comprises a convective term and a 
conduction one of the Fourier type. The resulting system of evolution
equations for particle number, momentum and energy, closed with the 
above constitutive law for the heat flux, form a parabolic system which is 
consistent with the entropy principle of linear irreversible thermodynamics
and the Onsager reciprocity relations. This system has been amply
investigated by various numerical techniques \cite{VRR}.

This parabolic system should be contrasted with
the extended hydrodynamical system of above, which is hyperbolic, and which
comprises evolution equations for particle number, momentum, energy and 
an evolution equation for the heat flux. For both systems the relaxation
times, which determine the transport coefficients and the production 
terms (arising from the collisions of electrons with the lattice phonons),
have been modeled in the same way and checked by Monte Carlo simulation
in a homogeneous slab of silicon. As a benchmark both systems have been
solved numerically for the submicron diode consisting of a 
$0.1 \; \mu m \;$ $n^{+}$ region followed by a 
$0.4 \; \mu m$ $n$ region, and ending with a $0.1 \; \mu m \;$
$n^{+}$ region. The doping density in the $n^{+}$ region 
is $N = 5 \times 10^7 \;
cm^{-3}$, while in the $n$ region it is 
$N = 2\times 10^{15} cm^{-3}$. The ambient device 
temperature is $T_{L} = 300$ Kelvin. 
We remark that the diode under consideration presents 
very steep density
gradients at the source and drain junctions and also 
very high electric fields
$( \geq 10^5 V/cm)$, which stretches up to its 
limits any continuum description.
Therefore this is an ideal benchmark for 
testing continuum models against
kinetic simulations.\\
The boundary conditions for the two models are consistent. The results for
the velocity profile 
have been compared with Monte Carlo simulations with exactly the same
conditions. It is apparent that the hyperbolic models, even in 
stationary situation, perform much better than the corresponding parabolic
one, vis a vis of the Monte Carlo simulations. Also for the velocity
profile there is a spike in the parabolic model near the second
junction, which is absent in the hyperbolic model (and also in
the ``true" Monte Carlo simulation -see Figure 2).
The spurious unphysical result
highlights the difference between the parabolic and 
hyperbolic model in regions of high electric fields 
and large gradients.
\end{enumerate}
\section{Relativistic examples}
Hyperbolic theories are conceptually more crucial 
in relativistics scenarios as  the ``action at a distance"
implied by parabolic theories finds no room there.  
\begin{enumerate}
\item {\it Precursor peak in astrophysical X-ray bursters}\\
Many fluids of astrophysical interest fall in the category of "radiative
plasmas". By such it is understand an interacting mixture of a material gas
(e.g. ionized hydrogen atoms plus electrons) and radiation quanta
(e.g. photons). The material medium is considered to be locally
in thermal equilibrium with itself as its internal
mean free times are extremely
short, whereas the radiation quanta, being out of equilibrium with the
material medium (its mean free time is finite) gives to  rise
dissipation. Obviously photons are scattered, 
absorbed and emitted by the gas \cite{MIHALAS}.
 
The analysis of elastic photon diffusion through a cloud
of ionized plasma of modest optical depth via parabolic
theories predicts for the distribution of the photons
over their time of escape from the cloud
the conventional diffusion structure,
but it is unable to account for the double peaked
temporal luminosity profiles observed from X-ray bursters. 
Hoffman {\em et al} \cite{HET} reported on a X-ray burster
showing a precursor peak, that lasted about 4 seconds,
neatly separated from the main event which lasted above 
1000 seconds. This pattern is reminiscent 
of those encountered 
in heat pulse experiments in solids alluded before,
it can be explained by the hyperbolic theory of photon
transport \cite{SCH}. The starting equations are
\[
\dot{N} + J^{a}_{,a} = S  
\]
and
\[
\kappa_{T} J^{a} + \dot{J}^{a} + \frac{1}{3} 
h^{ab} N_{,b} = 0
\]
with $ N^{a} ( = N u^{a} + J^{a})$
the photon four-current, $u^{a}$ the rest-frame
four-velocity of the material medium,
$N = -N^{a} u_{a}$
the photon number density, 
$J^{a} = h^{ab}N_{b}$ the convective photon
current relative to the ionized plasma,
$h^{ab} = g^{ab} + u^{a} u^{b}$ the 
spatial projector,
$S$ the photon source
(assumed Gaussian in time), and 
$\kappa_{T} = n \sigma_{T}$ where $n$ is the
electron number density and $\sigma_{T}$
the Thomson cross section.
The solution of the corresponding dispersion relation
is given in terms of a Green function showing
a diffusive behavior
for times longer than $t_{D} \sim \tau*^{2} t_{col}/2$
(the conventional diffusion time),
and a well defined peak for times shorter than $t_{D}$.
Here $\tau*$ stands for the optical depth. As seen in
figures 2, 3, 5, 6 and 7 of Ref \cite{SCH}
the solution is similar to that found in heat pulses
in solids.
\item {\it Thermal relaxation in gravitational collapse}\\
Very often in the analysis of the gravitational collapse 
of fluid spheres the thermal relaxation 
time $\tau$ is ignored \cite{BS} 
since it is usually very small as compared with the 
typical time scales of collapse for gravitating systems 
(for phonon-electron interaction $\tau \sim 10^{-11}$ sec,
and for phonon-phonon and free electron 
interaction $\tau \sim 10^{-13} \, \mbox{sec}$ at 
room temperature \cite{RP}). However, there are situations 
in which $\tau$ cannot be neglected against the
gravitational collapse. For instance, in cores of evolved
stars the quantum cells of phase space are filled up
and the electron mean free-path increases substantially
and so does $\tau$. For a completely degenerate core star
of radius one hundred of the solar radius and temperature 
of about ten million Kelvin one has 
$\tau \sim 1 \, \mbox{sec}$ \cite{MH}. 
In recent works Di Prisco {\em et al} \cite{PH} and
Herrera and Mart\'{\i}nez \cite{HM}
using the thermal conductivity by electrons in 
neutron star matter and the Maxwell-Cattaneo 
transport equation for the heat flux, 
showed that the outcome of the   
collapse of the star is very sensitive to the value of 
$\tau$ (they restricted themselves to the very conservative
range $ 10^{-6}\, \mbox{sec} \, < \tau < 10^{-3} \, \mbox{sec}$), 
to the point that a fluid sphere may
bounce if $\tau$ is shorter
than some critical value within the mentioned range,
otherwise it undergoes a complete gravitational 
collapse -see figure 4 of Ref. \cite{HM}. 
It is worthy of note that this 
analysis also applies (modulo some
obvious adaptation) to the Kelvin-Helmholtz phase
of the birth of a neutron star whose duration is 
about tens of seconds and in which most of the
binding energy (about $10^{53}$ erg) is emitted
\cite{BL}.

Further, phenomenae prior to relaxation
may influence in a important way the subsequent
evolution of the system (i.e., for times longer
than $\tau$). In this connection it has been shown 
that for spherically symmetric stars with a radial
heat flow, the temperature gradient appearing as 
a result of perturbations, and thereby the luminosity,
are highly dependent on the product of the 
relaxation time by the period of oscillation 
of the star \cite{LHN}.  
\item {\it Isotropic early Universe expasion}\\
It is natural to think that the fluid filling the
primeval Universe was subject to steep gradients,
whereby the very early cosmic world 
should be a right scenario for 
hyperbolic theories of dissipation to apply
together with Einstein's field
equations. As we will see by resorting to a 
couple of examples the
predictions of the causal theories greatly differ from
that of the acausal (parabolic) ones. Unfortunately
there is no compelling observational evidence favoring
one or another, as nowadays cosmic observation is  still
at its infancy.
As a first example let us consider a homogeneous and
isotropic universe with geometry described  by a
flat Robertson-Walker metric with scale factor $a(t)$.
And assume that the 
source of the metric is a barotropic fluid
(i.e.,  one with equation of state 
$P = (\gamma - 1) \rho \;$,
$ \gamma = $ constant $ >0$),
endowed with a nonvanishing bulk viscosity whose
coefficient depends on the energy density via
$\zeta = \alpha \rho^{m}$
with $\alpha$ a positive semidefinite constant and
$m$ a constant parameter. By using in this setting 
the parabolic Eckart's theory of irreversible 
phenomena Barrow \cite{BW} found 
the first order differential equation for the
evolution of the Hubble parameter
\[
2 \dot{H} + 3 (\gamma - 3^{m} \alpha H^{2m-1})H^{2} = 0 \; ,
\; \; (H \equiv \frac{\dot{a}}{a}) \; .
\]
For $m < 1/2$ this equation predicts, among other things, 
that  the Universe begins with a Friedmann
singularity and approaches a inflationary expansion
(i.e., $H = \; $ constant) 
for $t \rightarrow \infty$. However, if $m >1/2$
one has what it is termed as a ``deflationary" behavior,
i.e., the Universe starts with an inflationary expansion 
and evolves toward a Friedmann state. 

Things are different when a hyperbolic 
theory is used instead.
In such a case an expression for the relaxation time must
be postulated. Pav\'{o}n {\em et al} \cite{PBJ} choose
the well motivated relationship 
$\tau = \zeta/\rho \;$, later justified by 
Maartens \cite{RM}. The corresponding differential equation
for the evolution of the Hubble factor 
\begin{eqnarray}
\alpha \beta^{n}  H^{2n} \ddot{H} +  (1 + 3 \alpha \beta^{n}
\gamma H^{n+m}) \dot{H} \
&&\nonumber\\
+  \left( \frac{3}{2} \gamma -
\frac{12 \pi \alpha}{\beta^{m}} H^{n+m} \right) H^{2} = 0 &&\; ,
\nonumber
\end{eqnarray}
becomes second order
in time, and so the number of possible evolutions 
greatly augments. (Here $n \equiv m-1$, 
$\beta$ is a short hand for $3/(8 \pi)$, 
and units have been chosen so that 
$c = G = 1$). Last equation reduces to the former one under
the unphysical assumption of vanishing $\tau$, and in the very
particular case that  $\ddot{H} = -3\gamma H \dot{H}$.
The set of solutions of the causal equation is far richer
than that of the acausal equation, both for nearly 
stationary solutions and for the nonstationary ones
-see figures 1-5 in Ref \cite{PBJ}. Here we want to 
stress that the deflationary solutions found 
by Barrow \cite{BW}
for $m > 1/2$ are seen to be unstable. 
Initial inflationary
expansions can be found only by fine-tunining the initial 
conditions $H(0)$ and $\dot{H}(0)$, hence it is not a
general feature of cosmic evolution. Another point 
of interest refers to the 
``generalized second law of thermodynamics", which enters 
into play when the spacetime posses an event horizon.
In such cases is not the entropy of the cosmic fluid 
that musn't decrease with time ($\dot{S}_{f} \geq 0$), 
but the combined entropy of the event horizon $S_{H}$
and that of the fluid, i.e.,
$\dot{S}_{H} + \dot{S}_{f} \geq 0$. The conceptual
trouble with theories allowing unbounded
speeds for dissipative signals is that no meaningful
event horizon can be ascribed to the spacetime 
even though it experiences a de Sitter expansion.
So one is forced also in this regard 
to resort to hyperbolic (causal) theories if one wishes 
to deal with cosmic horizons when the 
source of the gravitational field is a dissipative fluid.
The concrete expressions for the generalized second law,
equation (3.53) of Ref \cite{BW} and 
equation (53) of Ref \cite{PBJ}, 
largely differ from one another. 
\item{\em Anisotropic early Universe expansion}\\
The above study has been extended to anisotropic spaces,
Bianchi-type I \cite{VD1} and Bianchi-type III \cite{VD2},
which admit shear stresses. In both instances the dynamical
equations turned out far too  complicated to allow analytical
integration, so qualitative analysis and numerical integration
were carried out instead. For Bianchi-type I 
the initial anisotropy dies away quickly and, in general, 
neither the Friedmann nor de Sitter expansions are stable.
Nevertheless, for a wide range of values of 
the parameters occurring in these equations, there exists
a stable submanifold of phase space where the latter
presents asymptotic stability. This contrasts with 
the findings of Huang \cite{HUA}, who used Eckart's
theory of dissipative processes [4.a] for the 
bulk and shear stresses, and concluded that 
the Bianchi-type I space asymptotically 
evolves either to
Friedmann or de Sitter states. 
For Bianchi-type III the Friedmann 
expansion is unstable and the
de Sitter expansion stable \cite{VD2}. 
\end{enumerate}
\section{Concluding remarks}
By resorting to some specific examples in non-relativistic
as well as in relativistic nonequilibrium thermodynamic
situations we have made explicit the fact that hyperbolic
(causal) theories of transport have a range of applicability 
much broader than the parabolic (acausal) ones. The differences
between both sets may be ignored in steady regimes only, i.e.,
when the relaxation times of the processes under consideration
are much shorter than the time scale of interest. This has
been known long since. It was originally hinted by Maxwell 
\cite{JCM}, and first seen in the laboratory by Peshkov 
\cite{PESH} and later confirmed many times over in variety
of experimental situations \cite{DDJ}, \cite{JCL}. 
In this way we have 
dispelled the claims recently made in \cite{RG} and \cite{LL} 
implying that hyperbolic theories have no 
useful applications. Note that in any case
the findings in  \cite{RG} and \cite{LL} are limited
to simple fluids and situations such that the quantity $\Gamma$,
defined above is much lower than unity. Thereby it does 
not cover many interesting physical fluid situations, such as
those recalled in this article. Therefore claims such as
{\it hyperbolic, relativistic theories for dissipative fluids,
reduce, physically, to Navier-Stokes} found in \cite{RG} cannot
be justified. In summary, the suggestion in \cite{RG} and
\cite{LL} that the class of solutions to which their theorems
apply include all solutions of the hyperbolic dissipative fluid
theories having physical relevance, is at variance with 
experimental evidence.

Current hyperbolic theories are not free, however, of some 
drawbacks, which will be mentioned in the following. This 
does not imply that one should go back to the acausal
parabolic theories, but instead try to improve on the existing
hyperbolic models. The main drawback of the moment methods
currently used in present day hyperbolic theories
is the introduction of spurious characteristic speeds, i.e.,
characteristic waves that arise only from the truncation 
procedure (the number of moments retained in the theory),
and have no counterpart in the underlying kinetic description.
Since these characteristics constitute a fan which broadens
as the number of moments increases, there could be 
serious difficulties (due to the mathematical properties 
of the hyperbolic systems) of the appearance of unphysical 
discontinuities in the solutions of the hyperbolic theories
at places where the fluid speed crosses one of the 
characteristic speeds \cite{RUGG}.
Fortunately the problem
is not so severe because lately Weiss \cite{WWS} has shown, 
by direct numerical integration (for the case of 
a rarefied gas of Maxwell molecules) that, by 
some algebraic cancellation, the discontinuities can occur
only at the highest characteristic speed, and the latter 
increases with the number of retained moments. If this 
result would be confirmed (and mathematically understood)
in more general cases, the aforesaid
difficulty would be alleviated.

More generally, the root of the difficulty lies 
in using the usual polynomial moments. In fact
an avenue for further research should be to 
investigate whether a different basis function would 
be more suitable. We remark that any basis function must
be able to capture phenomena varying on a very 
short time scale. In fact, the requirement of 
relativistic causality itself, forces the basis function
to be capable of describing high frequency situations. 
One formulation of relativistic causality is that the
wave speeds must be less or equal to the light speed 
in vacuum $c$. This must be applied to the 
characteristic speeds, but also to the dispersion waves
arising from a linearization of the equations. 
In this case the relativistic causality takes the form
$v_{g} \leq c \;$  as $\omega \rightarrow \infty \; $
(where $v_{g}$ is the group velocity),
and obviously this requires that the equations 
must be able to describe high frequency phenomena
(at least in the asymptotic sense). This is the main
physical reason why quasi-stationary theories,
based on slowly varying variables, are 
intrinsically unable to comply with relativistic causality
and makes it mandatory to consider hyperbolic theories.
\section{Acknowledgments}
The authors are indebted to David Jou, Luis Herrera and
Roy Maartens for conversations, constructive remarks 
and advice. This work was partially supported by the 
Spanish Ministry of Education under grant PB94-0718,  
MURST and CNR grants 95.01081.CT01 and 96.03855 CT01.

\newpage
\onecolumn
\hspace{2.cm}TABLE I
\[\]
\begin{tabular}{||l|c|c|c|c|c|c||}    \hline
$l/\lambda$           & 0.25 & 0.50 & 1.00 & 2.00 & 4.00 & 7.00 \\ \hline
$(c_s/v_{ph})_{FNS}$  & 0.40 & 0.26 & 0.19 & 0.13 & 0.10 & 0.07 \\
$(c_s/v_{ph})_{Hyp.}$ & 0.52 & 0.43 & 0.44 & 0.47 & 0.48 & 0.49 \\
$(c_s/v_{ph})_{Exp.}$ & 0.51 & 0.46 & 0.50 & 0.46 & 0.46 & 0.46 \\ \hline
\end{tabular}

\newpage
{\bf LIST OF CAPTIONS FOR TABLES}
\[\]
{\bf Table I}\\
Numerical values of $c_s/v_{ph}$ as a function of $l/\lambda$. 
\[\]
\[\]  
\[\]
\centerline{\bf LIST OF CAPTIONS FOR FIGURES}
\[\]
{\bf Figure 1}\\
Schematic results of heat pulse experiments. The dashed line 
represents the heat pulse at the source, whereas the solid
line shows the structure of the heat signal at an arbitrarily 
selected  point in in the sample.  Peaks {\bf A}, {\bf B} and  
{\bf C} correspond
to ballistic phonons with longitudinal sound velocity, ballistic
phonons with transverse sound speed, and second sound, respectively.
The tail {\bf D} corresponds to the diffusive heat propagation.
This structure can be explained by hyperbolic theories. By contrast,
parabolic theories are able to explain the tail {\bf D} only.
\[\]
{\bf Figure 2}\\
Velocity profile for the electrons in a submicron diode under
a very high electric field. The crosses correspond to the 
Monte Carlo simulation, the solid line to the prediction of
the hyperbolic theory, and the dashed line to the parabolic 
one. The latter shows a spurious peak. 

\begin{thebibliography}{99}
\bibitem{IS} W. Israel, {\it Ann. Phys. (N.Y.)} {\bf 100}, 310  (1976);
W. Israel and J. Stewart, {\it Ann. Phys. (N.Y.)} {\bf 118}, 341 (1979);
W. Israel, {\it Physica} A {\bf 106}, 204 (1981);
D. Pav\'{o}n, D. Jou and J. Casas-V\'{a}zquez, {\it Ann. Institute
H. Poincar\'{e}} A {\bf 36}, 79 (1982); 
I.S. Liu, I M\"{u}ller and T. Ruggeri, {\it Ann. Phys. (N.Y.)}
{\bf 169}, 191 (1986); B. Carter, in {\it Relativistic Fluid 
Dynamics}, edited by A.M. Anile and Y. Choquet-Bruhat
(Lectures Notes in Mathematics, Springer-Verlag, Berlin 1989);
R. Geroch and L. Lindblom, {\it 
Phys. Rev.} D {\bf 41}, 1855 (1990).
\bibitem{M} I. M\"{u}ller, {\it Arch. Rat. Mech.} 
{\bf 34}, 259 (1969);
I. M\"{u}ller and T. Ruggeri, {\it Extended Thermodynamics}
(Springer-Verlag, Berlin, 1993).
\bibitem{JCL} D. Jou, J. Casas-V\'{a}zquez and G. Lebon,
{\it Extended Irreversible Thermodynamics},
second edition
(Springer-Verlag, Berlin, 1996).
\bibitem{EL} C. Eckart, {\it Phys. Rev.} {\bf 58}, 919 (1940); 
L. Landau, {\it M\'{e}canique des Fluides} (MIR, Moscou, 1971).
\bibitem{HL} W. Hiscock and L. Lindblom, {\it Ann. Phys. (N.Y.)}
{\bf 151}, 466 (1983); {\it ibid.} {\it Phys. Rev.} D {\bf 31},
725 (1985).
\bibitem{RG} R. Geroch, {\it J. Math. Phys.} {\bf 36}, 4226 (1995).
\bibitem{LL} L. Lindblom, {\it Ann. Phys. (N.Y.)} {\bf 247}, 1 (1996).
\bibitem{JD} D. Jou and D. Pav\'{o}n, {\it Astrophys. J.}
{\bf 291}, 447 (1985).
\bibitem{PJC} D. Pav\'{o}n, D. Jou and J. Casas-V\'{a}zquez,
{\it J. Phys.} A {\bf 16}, 775 (1983).
\bibitem{ASS} See any standard textbook in Statistical Mechanics.
\bibitem{AMM} A.M. Anile and O. Muscato, {\it Phys. Rev.} B 
{\bf 51}, 16728 (1995).
\bibitem{WM} W. Weiss and I. M\"{u}ller, {\it Continuum 
Mech. Thermodyn.} {\bf 7}, 123 (1995).
\bibitem{JRD} J.R.D. Copley and S.W. Lovesey, 
{\it Rep. Progr. Phys.} {\bf 38}, 461 (1975).
\bibitem{MH} M. Harwit, {\it Astrophysical Concepts}
(Springer-Verlag, Berlin, 1988).
\bibitem{LEV} C.D. Levermore, {\it J. Stat. Phys.} {\bf 83}, 
1021 (1996).
\bibitem{GD} H. Grad, {\it Principles of the Kinetic Theory
of Gases}, Handbuch der Physik XII, edited by S. Flugge
(Springer-Verlag, Berlin, 1958).
\bibitem{PESH} V. Peshkov, {\it
J. Phys.} {\bf 8}, 381 (1944).
\bibitem{TZOU} D.Y. Tzou, {\it Macro-to Microscale Heat 
Transfer: The Lagging Behavior}, (Taylor \& Francis, 
Bristol, 1997) -chapter 4.
\bibitem{RAG}R.A. Guyer and J.A. Krumhansl, {\it
Phys. Rev.} {\bf 133}, 1411 (1964); {\it ibid} {\bf 
148} 778 (1966).
\bibitem{WDH} W. Dreyer and H. Struchtrup, {\it 
Continuum Mech. Thermodyn.} {\bf 5} 3 (1993).
\bibitem{DDJ} D.D. Joseph and L. Preziosi, {\it
Rev. Mod. Phys.} {\bf 61}, 41 (1989); {\it
ibid} {\bf 62}, 375 (1990).
\bibitem{AP} A.M. Anile and S. Pluchino, {\it J. M\'{e}canique}
{\bf 3}, 167 (1984).
\bibitem{LC} G. Lebon and A. Cloot, {\it Wave Motion} 
{\bf 11}, 23 (1989).
\bibitem{MS} E. Meyer and G. Sessler, {\it Z. f\"{u}r Phys.}
{\bf 149}, 15 (1975).
\bibitem{MG} M. Greenspan, in {\it Physical Acoustics},
vol. 2A, edited by W. P. Mason (Academic, New York, 1965).
\bibitem{CM} M. Carrassi and A. Morro, {\it
Nuovo Cimento} B {\bf 9}, 321 (1972). 
\bibitem{WM} W. Weiss and I. M\"{u}ller, {\it
Continuum Mech. Thermodyn.} {\bf 7}, 123 (1995).
\bibitem{CLARK} N.A. Clark, {\it
Physical Rev.} A {\bf 12}, 123 (1970).
\bibitem{KR} R. Kippenhahn, G. Ruschenplatt and H.C. Thomas,
{\it Astron. Astrophys.} {\bf 91}, 175 (1980). 
\bibitem{HF} L. Herrera and N. Falc\'{o}n, 
{\it Phys. Lett.} A {\bf 201}, 33 (1995).
\bibitem{BAH} R. B. Bird, R.C. Armstrong and O.H. Hassager,
{\it Dynamics of Polymeric Liquids}, second edition Vol 1
(J. Wiley, New York, 1987).
\bibitem{HAE} W. Haensch, {\it The Drift Diffusion Equation
and its Application in MOSFET Modelling} (Springer-Verlag,
Wien, 1991).
\bibitem{SEL} S. Selberherr, {\it Analysis and Simulation of
Semiconductor Devices} (Springer-Verlag, Berlin, 1994).
\bibitem{MRS} A. Markowich, C.A. Ringhofer and C. Schmeiser,
{\it Semiconductor Equations} (Springer-Verlag, Wien, 1990).
\bibitem{FLX} M.V. Fischetti, and S. Laux, 
{\it Phys. Rev.} B {\bf 48}, 2244 (1993).
\bibitem{APT} A.M. Anile, S. Pennisi and M.Trovato,
in {\it Nonlinear Hyperbolic Problems: Theoretical, Applied
and Computational Aspects}, edited by  A. Donato and 
F. Olivieri (Viveg, Stuttgart, 1993). 
\bibitem{AU} J.D. Au, {\it Nonlinear closure in molecular
extended thermodynamics with application to the shock 
wave structure solution} in Proceedings International Conference
on Wave and Stability, Palermo (1995).
\bibitem{AAT} A.M. Anile and M. Trovato, to be published in 
{\it Phys. Lett.} A (1997).
\bibitem{MUN} C. Truesdell and R.G. Muncaster, {\it
Fundamentals of Maxwell Kinetic Theory of Simple Monoatomic Gas}
(Pitman, Boston, 1985).
\bibitem{VRR} V. Romano and G. Russo, {\it
Numerical solutions for hydrodynamical models of semiconductors}
submitted to IEE trans. on CAD (1996);
A.M. Anile, C. Maccora and R.M. Pitadella, {\it
COMPEL} {\bf 14}, 1 (1995). 
\bibitem{MIHALAS} D. Mihalas and B.W. Mihalas {\it
Foundations of radiation hydrodynamics} (Oxford University 
Press, Oxford, 1984). 
\bibitem{HET} J.A. Hoffman, W.G.H. Lewin, J. Doty, J.G. Jernigan,
M. Haney and J.A. Richardson, {\it Astrophys. J. Lett.} 
{\bf 221}, L57 (1978).
\bibitem{SCH} A.M. Schweizer, {\it Astron. Astrophys.}
{\bf 151}, 79 (1985).
\bibitem{BS} W.B. Bonnor, A.K.G. Oliveira and N.O. Santos, 
{\it Phys. Rep.} {\bf 181}, 269 (1989).
\bibitem{RP} R. Peirls {\it Quantum Theory of Solids}
(Oxford University Press, London, 1956).
\bibitem{PH} A. Di Prisco, L. Herrera and M. Esculpi,
{\it Class. Quantum Grav.} {\bf 13}, 1053 (1996). 
\bibitem{HM} L. Herrera and J. Mart\'{\i}nez,  
{\it Gen. Relativ. Grav.} {\bf 30}, 445 (1998).
\bibitem{BL} A. Burrows and J. Lattimer, {\it
Astrophys. J.} {\bf 307}, 178 (1996).
\bibitem{LHN} L. Herrera and N.O. Santos,
{\it Mon. Not. R. Astr. Soc.} {\bf 287}, 161 (1997).
\bibitem{BW} J.D. Barrow, {\it Nucl. Phys.} B 
{\bf 310}, 743  (1988). 
\bibitem{PBJ} D. Pav\'{o}n, J. Bafaluy and D. Jou,
{\it Class. Quantum Grav.} {\bf 8}, 347 (1991). 
\bibitem{RM} R. Maartens, {\it Causal thermodynamics in 
relativity}, Lectures given at the Hanno Rund Workshop
on Relativity and Thermodynamics, University of 
Natal (1996), astro-ph/9609119.
\bibitem{VD1} V. Romano and D. Pav\'{o}n, {\it
Phys. Rev.} D {\bf 47}, 1396 (1993).
\bibitem{VD2} V. Romano and D. Pav\'{on}, {\it
Phys. Rev.} D {\bf 50}, 2572 (1994).
\bibitem{HUA}W.-H. Huang, {\it J. Math. Phys.}
{\bf 31}, 659 (1990); {\it ibid} {\bf 31}, 1456 (1990).
\bibitem{JCM} J. C. Maxwell, {\it Philos. Trans.
Soc. Lond.} {\bf 157}, 49 (1867).
\bibitem{RUGG} T. Ruggeri, {\it Phys. Rev.} E 
{\bf 47}, 4135 (1993).
\bibitem{WWS} W. Weiss, {\it Phys. Rev.} E {\bf 52},
R5760 (1995). 
\end{thebibliography}
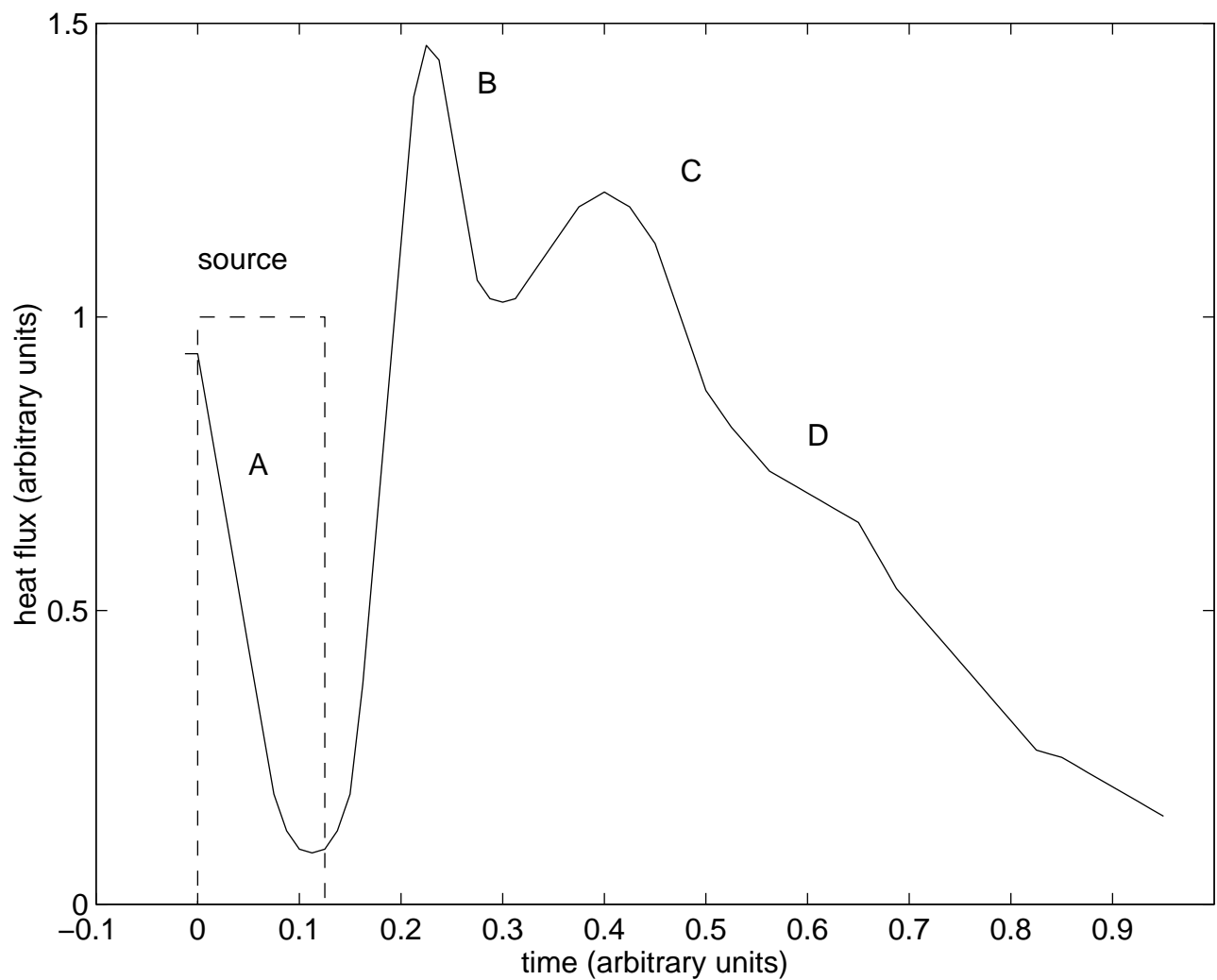
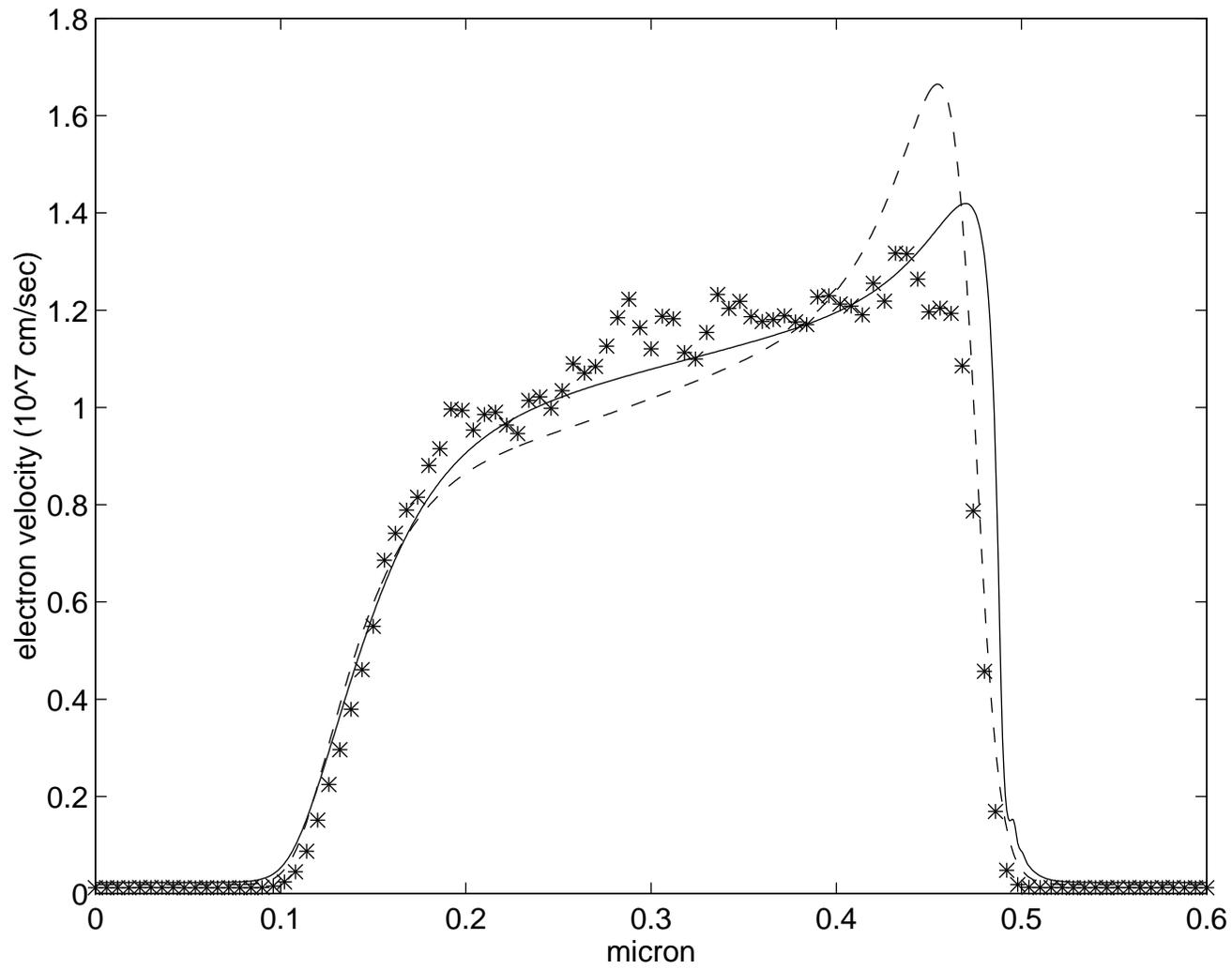
\end{document}